\documentclass[conference,dvi]{IEEEtranKV}

\usepackage{graphicx}
\usepackage{citesort}
\usepackage{amssymb}
\usepackage{amsfonts}
\usepackage{amsmath}
\usepackage{epsfig}
\usepackage{color}
\usepackage{fancybox}
\usepackage{textcomp}
\usepackage{multirow}
\usepackage{setspa ce}
\usepackage{psfrag}
\usepackage[ruled,vlined,linesnumbered]{algorithm2e}

\newtheorem{Proposition}{Proposition}

    \newcommand{\qh}{{\bf h}}

    \newcommand{\qx}{{\bf x}}
    
    \newcommand{\qz}{{\bf z}}

    \newcommand{\qH}{{\bf H}}
    \newcommand{\qI}{{\bf I}}

    \newcommand{\qQ}{{\bf Q}}

    \newcommand{\qU}{{\bf U}}
    
    \newcommand{\qW}{{\bf W}}
    \newcommand{\qX}{{\bf X}}
    \newcommand{\qY}{{\bf Y}}
    \newcommand{\qZ}{{\bf Z}}
    
    \newcommand{\qone}{{\bf 1}}

    \newcommand{\qSigma}{{\boldsymbol \Sigma}}

    \newcommand{\tc}{{\tilde{c}}}

    \newcommand{\tq}{{\tilde{q}}}

    \newcommand{\bbC}{{\mathbb C}}

    \newcommand{\calI}{{\mathcal I}}

    \newcommand{\calN}{{\mathcal N}}
    
    \newcommand{\calT}{{\mathcal T}}

    \newcommand{\calqH}{\boldsymbol{\cal H}}
    \newcommand{\calqQ}{\boldsymbol{\cal Q}}
    
    \newcommand{\calqX}{\boldsymbol{\cal X}}
    \newcommand{\calqZ}{\boldsymbol{\cal Z}}

    \newcommand{\tr}{{\sf tr}}
    \newcommand{\Ex}{{\sf E}}

    \newcommand{\Extr}{\operatornamewithlimits{\sf Extr}}
    
    \newcommand{\mse}{{\sf mse}}

    \newcommand{\rmd}{{\rm d}}
    
    \newcommand{\sfd}{{\sf d}}
    \newcommand{\sfj}{{\sf j}}
    \newcommand{\sft}{{\sf t}}

\title{Performance Limits of Massive MIMO Systems Based on Bayes-Optimal Inference}

\author{Chao-Kai Wen, Yongpeng Wu, Kai-Kit Wong, Robert Schober, and Pangan Ting
\thanks{C.-K. Wen is with the Institute of Communications Engineering, National Sun Yat-sen University, Kaohsiung, Taiwan (e-mail: $\rm ckwen@ieee.org$). Y. Wu and R. Schober are with the Institute for Digital Communications, Universit\"{a}t Erlangen-N\"{u}rnberg, Germany. K. Wong is with the Department of Electronic and Electrical Engineering, University College London, UK. P. Ting is with the Industrial Technology Research Institute, Hsinchu, Taiwan.}
}

\begin{document}
\maketitle
\begin{abstract}
This paper gives a replica analysis for the minimum mean square error (MSE) of a massive multiple-input multiple-output (MIMO) system by using Bayesian inference. The Bayes-optimal estimator is adopted to estimate the data symbols and the channels from a block of received signals in the spatial-temporal domain. We show that using the Bayes-optimal estimator, the interfering signals from adjacent cells can be separated from the received signals without pilot information. In addition, the MSEs with respect to the data symbols and the channels of the desired users decrease with the number of receive antennas and the number of data symbols, respectively. There are no residual interference terms that remain bounded away from zero as the numbers of receive antennas and data symbols approach infinity.
\end{abstract}

\section*{\sc I. Introduction}
Very large multiple-input multiple-output (MIMO) or ``massive MIMO'' systems \cite{Marzetta-10TW} are widely considered as a promising technology for future cellular networks. Given perfect channel state information (CSI) at the base station (BS), the array gain has been shown to grow unboundedly with the number of antennas at the BS so that both multiuser interference and thermal noise can be eliminated \cite{Larsson-14COMMag}. However, if the CSI is to be estimated using pilot signals, then the interference rejection capability is limited, even if the number of BS antennas is infinite. This phenomenon is known as {\em pilot contamination} \cite{Marzetta-10TW}.

Several promising approaches have recently emerged to mitigate pilot contamination \cite{Ashikhmin-12ISIT,Yin-13JSAC,Nguyen-13WCNC,Ngo-12ICASSP,Muller-14JSSP,Ma-14TSP}. The schemes in \cite{Ashikhmin-12ISIT,Yin-13JSAC,Nguyen-13WCNC} exploit a specific structure of the channels in the spatial domain while \cite{Ngo-12ICASSP,Muller-14JSSP} propose a singular value decomposition (SVD) based blind channel estimation scheme which projects the received signals onto an (almost) interference-free subspace to overcome pilot contamination. Recently, however, in \cite{Ma-14TSP} a the data-aided scheme, which jointly estimates the channels and data, was shown to be more effective then the schemes in \cite{Ngo-12ICASSP,Muller-14JSSP}. Hence, it is of great interest to study the best possible performance of massive MIMO systems employing joint channel-and-data (JCD) estimation.

To address this issue, unlike other JCD estimation schemes based on suboptimal criteria \cite{Takeuchi-13TIT,Ma-14TSP}, we use Bayes-optimal inference for JCD estimation as this approach provides the minimal mean-square-error (MSE) with respect to (w.r.t.) the channels and the payload data. However, an implementation of the Bayes-optimal JCD estimator appears extremely challenging not to mention analysing the resulting performance. In this paper, using the replica method from statistical physics, we are able to provide analytical results for the MSEs w.r.t.~the channels and the payload data for the Bayes-optimal JCD estimator. The quest for a practical implementation of the estimator is left for future study. Nevertheless, the computed MSE constitutes a useful lower bound for any suboptimal estimation scheme.
The mathematical framework we use herein is similar to that for solving the matrix factorization problem in \cite{Krzakala-13ISIT,Sakata-13ISIT,Kabashima-14ArXiv} but the adopted tools and derivations are adapted to the massive MIMO context. Several key observations are made from the analysis:
\begin{enumerate}
\item With the Bayes-optimal estimator, data can be estimated reliably even with negligible pilot information and the interference from adjacent cells can be separated from the received signals without pilot information.
\item The MSEs w.r.t.~the data symbols and the channels decrease with the number of receive antennas $N$ and the number of data symbols $T$, respectively. In addition, residual interference terms tend to zero as $N,T \to \infty$.
\item Since $T$ is limited by the channel coherence time, a non-zero MSE w.r.t.~the channel matrix remains as $N\to\infty$.
\end{enumerate}

\section*{\sc II. System Model}
\subsection*{A. Channel Model}
We consider an uplink wireless communication system with $C$ cells, where each cell contains a BS and there are $K_c$ user equipments (UEs) in cell $c$. Each BS
has $N$ antennas and each UE has one antenna. We assume that the channel is constant for some discrete time interval $T$ (in symbols). For ease of exposition, we
let the first cell be the cell of interest. The transmit symbols in the $c$th cell can be represented by matrix $\qX_{c} \in \bbC^{K_c \times T}$, and the
corresponding channel vector between the UEs in the $c$th cell and the target BS is denoted by $\qH_c \in \bbC^{N \times K_c}$. The received signals within one
block of $T$ symbols at the target BS can be written in matrix form as
\begin{equation} \label{eq:sys}
\qY = \frac{1}{\sqrt{K}} \sum_{c=1}^{C}\qH_c\qX_c + \qW \triangleq \frac{1}{\sqrt{K}}\qH \qX + \qW,
\end{equation}
where $\qW \in \bbC^{N \times T}$ denotes the temporally and spatially white Gaussian noise with zero mean and element-wise variance $\sigma^2$. Also, in
(\ref{eq:sys}), we have defined $\qX \triangleq [ \qX_1^{\dag} \cdots \qX_C^{\dag} ]^{\dag} \in \bbC^{K \times T}$ and $\qH \triangleq \left[ \qH_1 \cdots
\qH_C\right] \in \bbC^{N \times K}$ with $K = \sum_{c=1}^{C} K_c$. Here, $(\cdot)^{\dag}$ denotes the Hermitian conjugate operation.

\begin{figure}
\begin{center}
\resizebox{3.0in}{!}{%
\includegraphics*{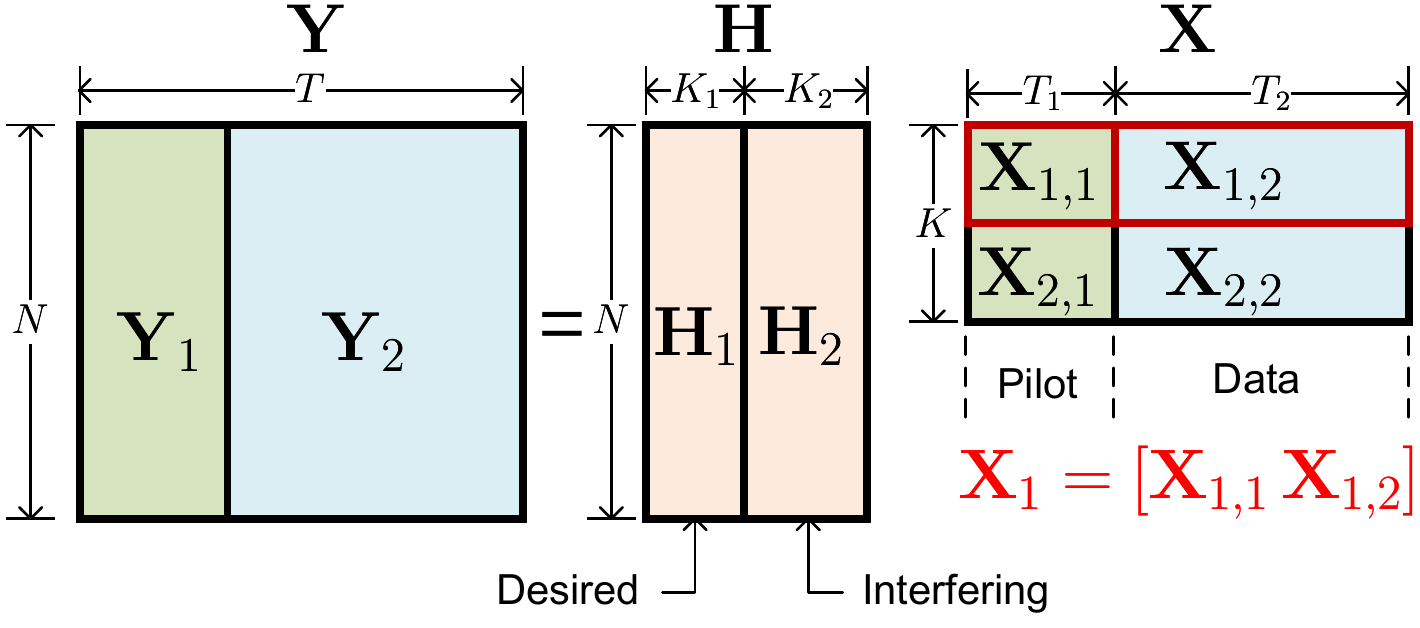} }%
\caption{An example of the received signal structure.}\label{fig:sysBlock}
\end{center}
\end{figure}

Since $\qH$ is not known to the receiver, a pilot-aided transmission is adopted. In particular, within one block of $T$ symbols, all UEs in the $C$ cells
simultaneously transmit \emph{different} pilot sequences of length $T_{1}$ symbols for channel estimation, and the remaining $T_2 = T-T_1$ symbols are used for
data transmission. The training and data phases, referred to as $\sft$-phase and $\sfd$-phase, respectively, are equivalent to partitioning $\qX_c$ as $\qX_c =
[\qX_{c,1} \, \qX_{c,2}]$ with $\qX_{c,1}  \in \bbC^{K_c \times T_1}$ and $\qX_{c,2}  \in \bbC^{K_c \times T_2}$. Also, $\qY$ is partitioned as $\qY = [\qY_1 \,
\qY_2]$ with $\qY_1  \in \bbC^{N \times T_1}$ and $\qY_2  \in \bbC^{N \times T_2}$. Figure \ref{fig:sysBlock} shows an example of the received signal structure for
$C=2$.

We assume that $\qX_{c,1} (\qX_{c,2})$ is composed of independent and identically distributed (i.i.d.)~random variables $X_{c,1}^{ij} (X_{c,2}^{ij})$ generated
from a probability distribution $P_{X_{c,1}} (P_{X_{c,2}})$, i.e.,
\begin{equation} \label{eq:ProX}
 P_X(\qX) = \prod_{c=1}^{C} P_{X_{c,1}}(\qX_{c,1})P_{X_{c,2}}(\qX_{c,2})
\end{equation}
with $P_{X_{c,1}}(\qX_{c,1}) = \prod_{i,j} P_{X_{c,1}}(X_{c,1}^{ij})$ and $P_{X_{c,2}}(\qX_{c,2})= \prod_{i,j} P_{X_{c,2}}(X_{c,2}^{ij})$. The transmit powers during $\sft$-phase and
$\sfd$-phase are denoted by $\Gamma_1$ and $\Gamma_2$, respectively. Hence, $\Ex\{ |X_{c,t}^{ij}|^2\} = \Gamma_t, \forall c$, where $\Ex\{ \cdot \}$ denotes
statistical expectation.

Similarly, we assume that $\qH_c$ is composed of i.i.d.~random variables $H_c^{ij}$ drawn from a zero-mean complex Gaussian distribution with variance $G_c$, i.e.,
$H_c^{ij} \sim P_{H_c} \equiv \calN_{\bbC}(0,G_c)$. Here, $G_c$ is the large-scale fading factor for the UEs in the $c$th cell to the target BS.\footnote{Our
results can be easily extended to the case where the UEs in one cell have different large-scale fading factors. The assumption that all UEs in a cell have the same
large-scale fading factor is made to simplify the notation.} As a result, we have
\begin{equation} \label{eq:ProH}
 P_H(\qH) = \prod_{c=1}^{C} P_{H_c}(\qH_c)
\end{equation}
with $P_{H_c}(\qH_c) = \prod_{i=1}^{N} \prod_{j=1}^{K_c} P_{H_c}(H_c^{ij}) $.

\subsection*{B. Bayes-Optimal Estimator}
We assume that neither the UEs nor the BS have CSI. Only the statistical properties of the channel and the transmit symbols as well as the pilot symbols are assumed known at the target BS. The
objective of the BS receiver is to estimate both $\qH_1$ and $\qX_{1,2}$ from $\qY$ given $\qX_{1,1}$. We employ the Bayes-optimal estimator which achieves the
minimum MSE (MMSE) w.r.t. channel $\qH_1$ and data $\qX_{1,2}$.

From (\ref{eq:sys}), the conditional distribution of $\qY$ given $(\qH,\qX)$ is
\begin{equation} \label{eq:ProYcondHX}
    P(\qY|\qH,\qX) = \frac{1}{(\pi \sigma^2)^{NT}} e^{-\frac{1}{\sigma^2} \| \qY - \frac{1}{\sqrt{K}}\qH\qX \|^2},
\end{equation}
where $\| \cdot \|$ denotes the Frobenius norm. Following the Bayes theorem, the posterior distribution of $\qH$ and $\qX$ given $\qY$ is
\begin{equation} \label{eq:posteriorDis}
    P(\qH,\qX|\qY) = \frac{P_H(\qH) P_X(\qX) P(\qY|\qH,\qX)}{Z(\qY)},
\end{equation}
where $Z(\qY)$ is a normalization constant. The Bayes-optimal estimate for $X_{c,t}^{ij} $  for $c=1, \dots,C$ and $t=1,2$ is therefore given by \cite{Poor-94BOOK}
\begin{equation} \label{eq:estx}
    \hat{X}_{c,t}^{ij} = \Ex\{X_{c,t}^{ij}|\qY\} = \int X_{c,t}^{ij} \, \phi_X(X_{c,t}^{ij}) \rmd X_{c,t}^{ij} ,
\end{equation}
where $\phi_{X}(X_{c,t}^{ij})$ is the marginal probability of $X_{c,t}^{ij}$ w.r.t.~the posterior distribution $P(\qH,\qX|\qY)$, i.e.,
\begin{equation}
\phi_X(X_{c,t}^{ij}) =\int_{\qH} \int_{\qX \backslash X_{c,t}^{ij}} P(\qH,\qX|\qY) \rmd \qX \rmd \qH,
\end{equation}
where the notation $\int_{\qX \backslash X_{c,t}^{ij}} \rmd \qX$ denotes the integration over all variables in $\qX$ except for $X_{c,t}^{ij}$. Similarly, the
Bayes-optimal estimate of $H_{c}^{ij} $ is given by
\begin{equation} \label{eq:estH}
    \hat{H}_{c}^{ij} = \Ex\{H_{c}^{ij}|\qY\} = \int H_{c}^{ij} \, \phi_H(H_{c}^{ij}) \rmd H_{c}^{ij}
\end{equation}
with $\phi_H(H_{c}^{ij}) = \int_{\qH\backslash H_{c}^{ij}} \int_{\qX} P(\qH,\qX|\qY) \rmd \qX \rmd \qH $.

In fact, the objective of the Bayes-optimal estimator is to approximate the received signal matrix $\qY$ by a rank $K$ product $\frac{1}{\sqrt{K}}\qH \qX$, where
$\qH$ is a tall matrix and $\qX$ is a wide matrix with the constraint on its upper left (corner) submatrix being $\qX_{1,1}$.

Defining $\qH_{\calI} \triangleq\left[ \qH_2 \cdots \qH_C \right] $ and $\qX_{\calI} \triangleq [ \qX_2^{\dag} \cdots \qX_C^{\dag} ]^{\dag} $, we can rewrite
(\ref{eq:sys}) as $\qY = \qZ_1 + \qZ_{\calI}  +\qW$ with $\qZ_1 \triangleq \frac{1}{\sqrt{K}} \qH_1\qX_1$ and $\qZ_{\calI} \triangleq \frac{1}{\sqrt{K}}
\qH_{\calI}\qX_{\calI}$. Since the pilot information of the inter-cell interference is not available, it is only possible for the estimator to obtain $\qH_{\calI}
\qU$ and $\qU^{-1} \qX_{\calI}$ where $\qU$ is any nonsingular matrix. Fortunately, this ambiguity does not change $\qZ_{\calI} = \frac{1}{\sqrt{K}}\qH_{\calI}
\qX_{\calI} $. Thus, with the Bayes-optimal estimator, $\qZ_{\calI}$ can be separated from $\qY$ without pilot information. On the other hand, the pilot
information $\qX_{1,1}$ can be used to eliminate the ambiguity regarding $\qH_1$ and $\qX_{1,2}$.

\textit{Remark 1:} The Bayes-optimal estimator is closely related to the principle component analysis (PCA). However, the Bayes-optimal estimator can exploit additional structural information (e.g., $\qX_{1,1}$) about the principal eigenvectors. In this context, let us investigate the relation between the Bayes-optimal estimator and the SVD-based blind channel estimation scheme \cite{Ngo-12ICASSP,Muller-14JSSP} which is also closely related to PCA. In the SVD-based scheme, the
interfering signal $\qZ_{\calI}$ is mitigated by projecting $\qY$ onto an (almost) interference-free subspace, obtained from the SVD of $\qY$.
If $N$ is very large and there is a certain power margin between $G_1$ and $G_\calI \triangleq \{ G_2,\dots, G_C\}$, the first $K_1$ columns of the left singular
vectors of $\qY$ are highly correlated with $\qH_1$ and also almost orthogonal to $\qH_{\calI}$. However, the pilot information $\qX_{1,1}$ is not exploited by the
SVD-based scheme at this stage. Hence, a power margin between $G_1$ and $G_\calI$ is needed to separate the signal subspace from the interference subspace. In contrast, the Bayes-optimal estimator does not have such limitation as it incorporates the knowledge of $\qX_{1,1}$.

In this paper, our aim is to derive analytical results for the MSEs of $\qX_{1,2}$ and $\qH_1$ for the Bayes-optimal
estimator. 
The average MSEs w.r.t. $\qX_{c,t}$ and $\qH_c$ are defined as
\begin{align}
    &\mse_{X_{c,t}} \triangleq \frac{1}{K_c T_t} {\int_{\qH} \int_{\qX} \|\hat{\qX}_{c,t} - \qX_{c,t}\|^2 P(\qH,\qX|\qY) \rmd \qX \rmd \qH}, \label{eq:mseX} \\
    &\mse_{H_c} \triangleq \frac{1}{K_c N} {\int_{\qH} \int_{\qX} \|\hat{\qH}_{c} - \qH_{c}\|^2  P(\qH,\qX|\qY) \rmd \qX \rmd \qH}. \label{eq:mseH}
\end{align}

\section*{\sc IV. MSE Analysis}
\subsection*{A. Scalar Channel}
First, we review the Bayes-optimal estimator (\ref{eq:estx}) for scalar channel
\begin{equation} \label{eq:ScalarSysModel}
    Y_{\tq_X} = \sqrt{\tq_X} X + W,
\end{equation}
where $X$ and $W \sim \calN_{\bbC}(0,1)$ are independent. This is a special case of (\ref{eq:sys}) with $M=N=T=1$. Following (\ref{eq:estx}), the MMSE estimate of
$X$ from $Y_{\tq_X}$ is given by
\begin{equation} \label{eq:ScalarEstX}
    \hat{X} = \int X \phi(X|Y_{\tq_X}) \rmd X \triangleq \Ex\{ X | Y_{\tq_X} \},~\mbox{where}
\end{equation}
$\phi(X|Y_{\tq_X}) = \frac{P(Y_{\tq_X}|X) P(X)}{P(Y_{\tq_X})}$, $P(Y_{\tq_X}|X) =\frac{e^{- |Y_{\tq_X}-\sqrt{\tq_X} X|^2}}{\pi}$. Note that $\hat{X}$ depends on $Y_{\tq_X}$ while we have suppressed $Y_{\tq_X}$ for brevity. Then, from (\ref{eq:mseX}), the MSE w.r.t.~$X$ is
\begin{equation} \label{eq:defMMSE}
    \mse_{X} = \Ex\{|X- \Ex\{ X | Y_{\tq_X} \} |^2\},
\end{equation}
where the expectation is taken over the joint conditional distribution $P(Y_{\tq_X},X) = P(Y_{\tq_X}|X) P(X)$.

Explicit expressions for the MSE w.r.t. $X$ are available for some special distributions. For example, if $X \sim \calN_{\bbC}(0,\Gamma)$, we have $\hat{X} =
\frac{\Gamma\sqrt{\tq_X}}{1+\Gamma \tq_X} Y_{\tq_X}$ which leads to $\mse_{X} = \frac{\Gamma}{1+\Gamma\tq_X}$.
Alternatively, if the signal is drawn from a quadrature phase shift keying (QPSK) constellation, the MSE w.r.t.~$X$ is given by \cite[Eq. (26)]{Lozano-06TIT}.

\subsection*{B. Massive MIMO}
Although an analytical expression for the MSE for the scalar channel is available, the task of developing the corresponding result for the massive MIMO channel
(\ref{eq:sys}) appears challenging. Existing efforts \cite{Guo-05IT,Bayati-11IT} show such development, if $\qH$ is perfectly
known at the BS receiver.

Our interest, however, is the case when $\qH$ is unknown. Our analysis studies the high-dimensional regime where $N, K, T \to \infty$ but the ratios
$N/K= \alpha$, $T/K = \beta$, $T_t/K =  \beta_t$, for $t=1,2$, and $K_c/K = k_{c}$, for $c=1,\dots,C$ are fixed and finite. For convenience, we simply use $K \rightarrow \infty$ to
denote this high-dimension limit. Following the argument of \cite{Krzakala-13ISIT,Kabashima-14ArXiv}, it can be shown that $\mse_{X_{c,t}}$ and $\mse_{H_c}$
are saddle points of the average free entropy
\begin{equation}\label{eq:FreeEn}
    \Phi \triangleq \frac{1}{K^2}\Ex_{\qY}\left\{\log Z(\qY)\right\},
\end{equation}
where $Z(\qY)$ is the normalization constant in (\ref{eq:posteriorDis}) given by
\begin{equation} \label{eq:partFun}
Z(\qY)={\int \rmd \qX P(\qX) \int \rmd \qH P(\qH) \, \frac{e^{-\frac{1}{\sigma^2}\left\| \qY - \frac{1}{\sqrt{K}} \qH \qX \right\|^2}}{(\pi\sigma^2)^{NT}}}.
\end{equation}
The major difficulty in computing (\ref{eq:FreeEn}) is the expectation over $\qY$. We can, nevertheless, facilitate the mathematical derivation by rewriting $\Phi$
as \cite{Nishimori-01BOOK}
\begin{equation}\label{eq:LimF}
\Phi = \frac{1}{K^2} \lim_{\tau\rightarrow 0}\frac{\partial}{\partial \tau}\log \Ex_{\qY}\left\{Z^{\tau}(\qY)\right\},
\end{equation}
where we have moved the expectation operator inside the log-function. We first evaluate $\Ex_{\qY}\left[Z^{\tau}(\qY)\right]$ for an integer-valued $\tau$, and
then generalize the result to any positive real number $\tau$. This technique is called the replica method, and has been widely adopted in the statistical physics
\cite{Nishimori-01BOOK} and information/communications theory literature, see e.g.,
\cite{Tanaka-02IT,Moustakas-03TIT,Guo-05IT,Muller-03TSP,Hatabu-09PRE,Takeuchi-13TIT,Girnyk-14TWC}. Under the assumption of replica symmetry, the following results
are obtained.

\begin{Proposition} \label{Pro_MSE}
As $K \to \infty$, the asymptotic MSEs w.r.t.~$\qX_{c,t}$ and $\qH_c$ are given by
\begin{align} \label{eq:asyMSE}
 \mse_{H_c} &= \Ex\{|H_c - \Ex\{ H_c | Y_{\tq_{H_c}} \} |^2\}, \\
 \mse_{X_{c,t}} &= \Ex\{|X_{c,t}- \Ex\{ X_{c,t} | Y_{\tq_{{X_{c,t}}}} \}|^2\},
\end{align}
where $Y_{\tq_{H_c}} \triangleq \sqrt{\tq_{H_c}} H_c + W_{H}$ and $Y_{\tq_{{X_{c,t}}}} \triangleq \sqrt{\tq_{{X_{c,t}}}} X_{c,t} + W_{X}$ with $W_{H},W_{X} \sim
\calN_{\bbC}(0,1)$ being independent of $H_c \sim P_{H_c}$ and $X_{c,t} \sim  P_{X_{c,t}}$, and
\begin{align}
  \tq_{H_c} &= \sum_{t=1}^{2}  \frac{\beta_t (c_{X_{c,t}}-\mse_{X_{c,t}}) }{\sigma^2+\sum_{l=1}^{C} k_{l} \Delta_{l,t} }, \label{eq:asyVarH} \\
  \tq_{X_{c,t}} &= \frac{\alpha (c_{H_c}-\mse_{H_c})}{\sigma^2+\sum_{l=1}^{C} k_{l} \Delta_{l,t}}, \label{eq:asyVarX}
\end{align}
with $\Delta_{c,t} = \mse_{H_c} c_{X_{c,t}} + \mse_{X_{c,t}} (c_{H_c}-\mse_{H_c})$, $c_{X_{c,t}} = \Ex\{|X_{c,t}|^2\} = \Gamma_t$, and $c_{H_{c}} = \Ex\{|H_{c}|^2\} =
G_c$.
\end{Proposition}
\hspace{-0.35cm}
\begin{proof}
A sketch of the proof is given in the appendix.
\end{proof}

We consider a few practical special cases of Proposition \ref{Pro_MSE} in the next section.

\section*{\sc V. Discussions and Numerical Results}
{\bf Example 1 (Single Cell and Perfect CSI)}---Let us begin by investigating the simplest case where there is only a single cell, i.e., $C=1$, and the channel matrix $\qH$ is perfectly known. In this case, the $\sft$-phase is not required so $\beta_2 = \beta$ and $\beta_1 = 0$. Since there is only one cell and only one phase in $\qX$, we omit the cell and phase indices $(c,t)$ from all the concerned parameters in this example. Because $\qH$ is perfectly known, we set $\mse_{H} =0$. Plugging these parameters into (\ref{eq:asyVarX}), we obtain
\begin{equation} \label{eq:effNoiseLevelX_PerfCh}
  \tq_X = \frac{\alpha G }{\sigma^2 + G \mse_{X} }
\end{equation}
with $\mse_{X} = \Ex\{ |X-\Ex\{ X | Y_{\tq_{X}} \}|^2 \} $. This MSE expression agrees with \cite[Eq. (1.6)]{Bayati-11IT}.

{\bf Example 2 (Single Cell and Unknown CSI)}---With only one cell, we omit the index $c$ from all the concerned parameters. In the conventional pilot-only scheme, the receiver first uses $\qY_1$ and $\qX_1$ to generate an estimate of channel $\qH$ and then uses the estimated channel for estimating the data $\qX_2$ from $\qY_2$. In this case, the MSE w.r.t.~$\qH$ under the Bayes-optimal approach (i.e., the MMSE estimates) can be obtained from Proposition \ref{Pro_MSE}. In fact, the analysis is the same as that in Example 1 but the roles of $\qH$ and $\qX_1$ are exchanged. Specifically, in the $\sft$-phase, i.e., $t=1$, the pilot matrix $\qX_1$ is known. Therefore, substituting $\mse_{X_1} = 0$ into (\ref{eq:asyVarH}) yields
\begin{equation} \label{eq:effNoiseLevelH_SinglCell_TrainPhase}
  \tq_H = \frac{\beta_1 \Gamma_1}{\sigma^2+ \Gamma_1 \mse_{H}} ~\mbox{ and }~ \mse_{H} = \frac{G}{1+G\tq_H}.
\end{equation}
Note that the MSE in (\ref{eq:effNoiseLevelH_SinglCell_TrainPhase}) is obtained under the assumption that $\qX_1$ is drawn from a zero-mean i.i.d.~sequence. In
addition, we have exploited the fact that each entry of $\qH$ is drawn from the complex Gaussian distribution.

\begin{figure}
\begin{center}
\resizebox{3.5in}{!}{%
\includegraphics*{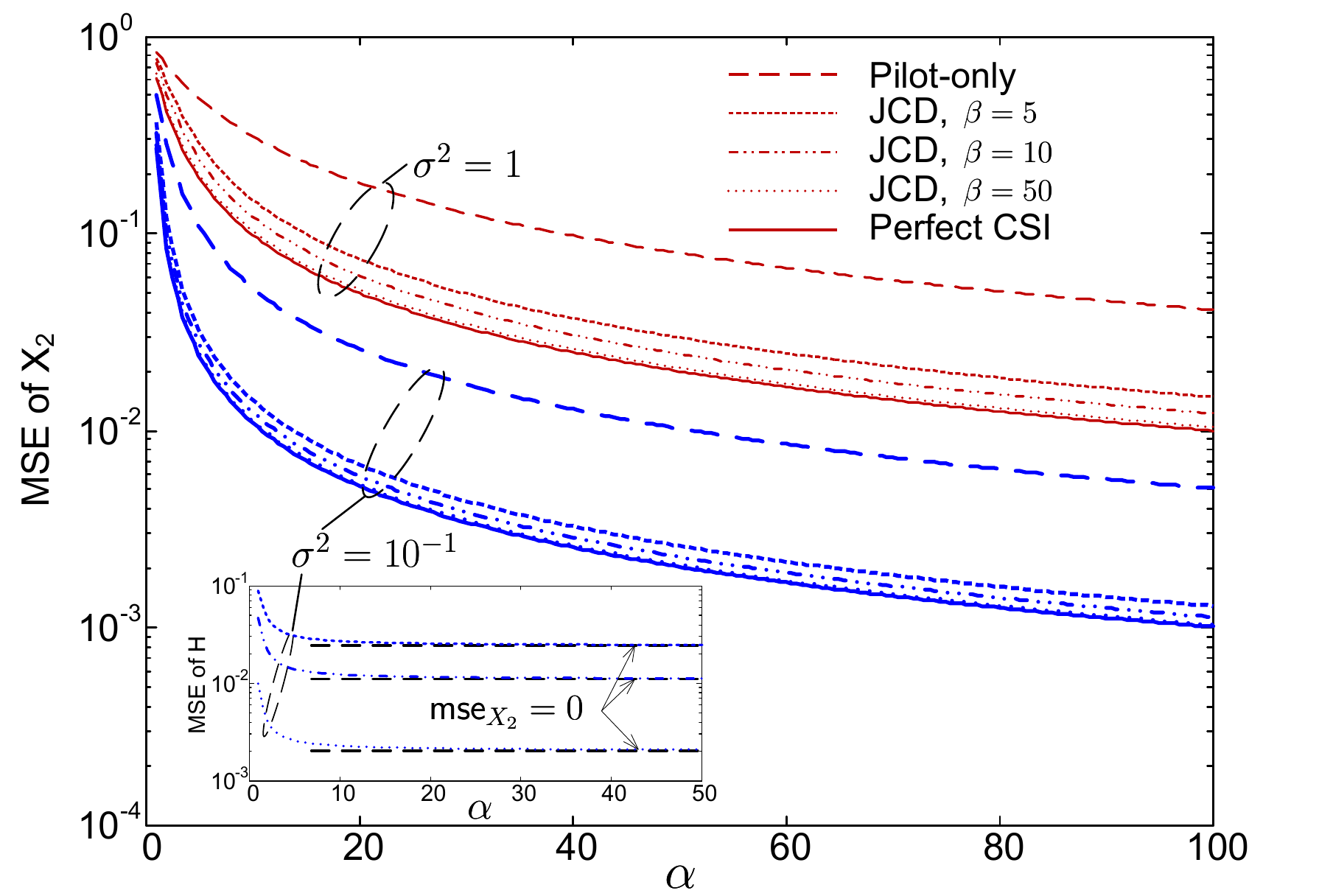} }%
\caption{The MSEs w.r.t.~$\qX_2$ for different estimation schemes in the single cell setup. $\beta_1=1$, $\Gamma=G=1$, and $X^{ij} \sim \calN_{\bbC}(0,\Gamma)$.}\label{fig:MSEX_All}
\end{center}
\end{figure}

\begin{figure}
\begin{center}
\resizebox{3.5in}{!}{%
\includegraphics*{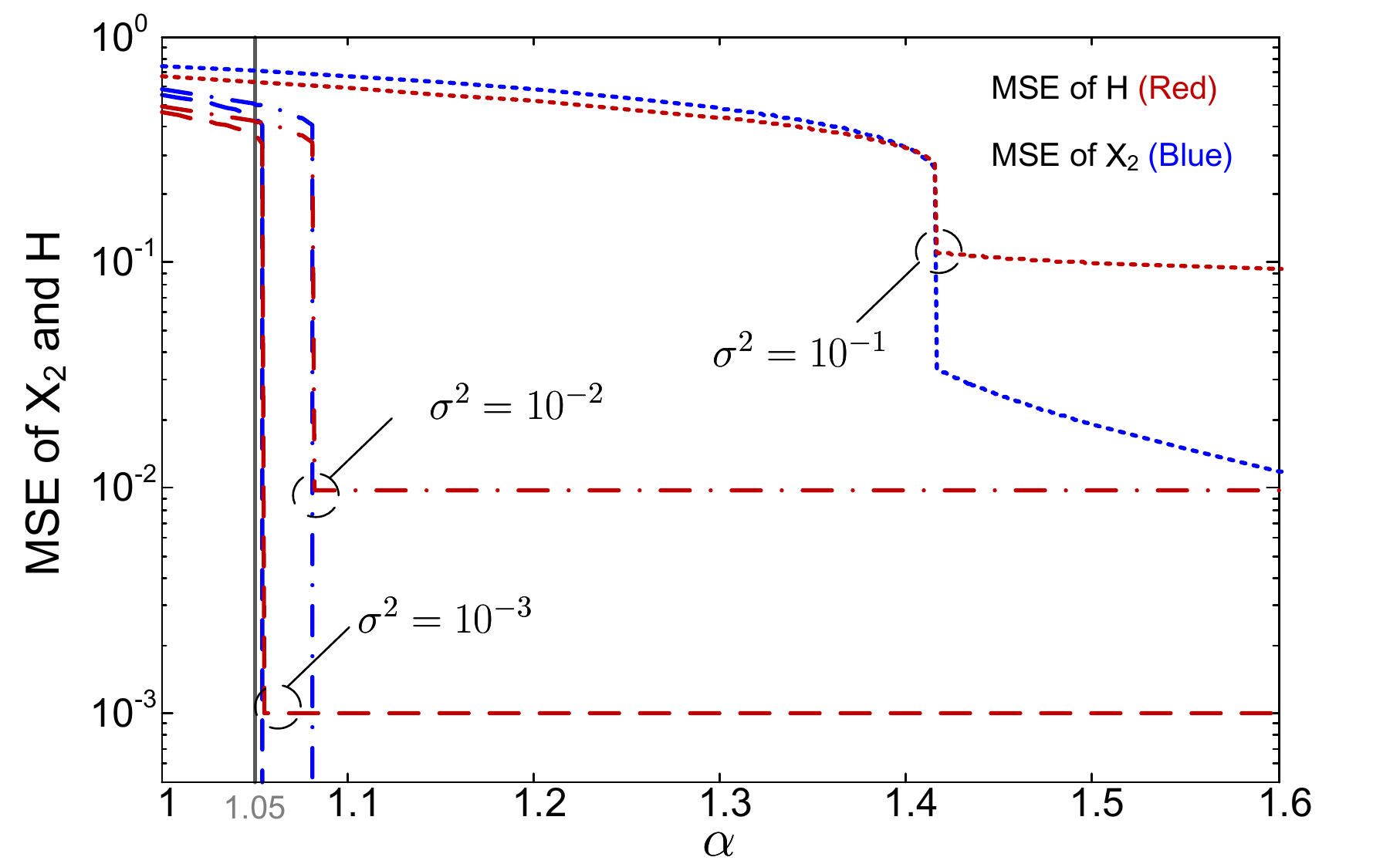} }%
\caption{The MSEs w.r.t. $\qX_2$ and $\qH$ as functions of $\alpha$ for the JCD estimation scheme. $\beta=2$, $\beta_1=10^{-4}$, $\Gamma=G=1$, and $P(X^{ij}) = \frac{1}{4} \delta( X^{ij} \pm 1/\sqrt{2} \pm \sfj/\sqrt{2} ) $. The vertical full line (in gray) marks the transition for $\sigma^2 = 0$. }\label{fig:MSEXH_DataAid_QPSKANoiseless}
\end{center}
\end{figure}

If the channel estimate along with its error statistic is used for data estimation under the Bayes-optimal approach, we can get the MSE w.r.t.~$\qX_2$ by applying the methods in \cite{Krzakala-13ICASSP,Wen-10TWC}
\begin{equation} \label{eq:effNoiseLevelX_SinglCell_TrainPhase}
  \tq_{X_2} = \frac{\alpha (G-\mse_{H})}{\sigma^2 + G \mse_{X_2} +  \mse_{H}(\Gamma_2-\mse_{X_2})},
\end{equation}
where $\mse_{X_2} = \Ex\{ |X_2- \Ex\{ X_2|Y_{\tq_{X_2}} \} |^2 \} $, and $\mse_{H}$ is given by (\ref{eq:effNoiseLevelH_SinglCell_TrainPhase}).
Recall that $\tq_{X_2}$ represents the \emph{effective} signal-to-interference-plus-noise ratio (SINR) of the equivalent scalar channel in
(\ref{eq:ScalarSysModel}), which measures the quality of the estimate of $\qX_2$. Comparing (\ref{eq:effNoiseLevelX_SinglCell_TrainPhase}) with
(\ref{eq:effNoiseLevelX_PerfCh}), we observe that the channel estimation error $\mse_{H} \neq 0$ not only results in a reduction of the numerator, as $\alpha$ is
multiplied by $(G-\mse_{H})$, but also an increase in the denominator through $\mse_{H}(\Gamma_2-\mse_{X_2})$.

Next, we study the case with JCD estimation. Specifically, we estimate both $\qH$ and $\qX_2$ from $\qY$ given $\qX_1$ using Bayes. Because the pilot matrix
$\qX_1$ is known, we substitute $\mse_{X_1} = 0$ into (\ref{eq:asyVarH}). Then, we obtain
\begin{align}
  \tq_H &= \frac{\beta_1  \Gamma_1}{\sigma^2+ \Gamma_1 \mse_{H}} + \frac{\beta_2 (\Gamma_2-\mse_{X_2})}{\sigma^2 + G \mse_{X_2} +  \mse_{H}(\Gamma_2-\mse_{X_2})}. \label{eq:effNoiseLevelH_SinglCell}
\end{align}
Comparing (\ref{eq:effNoiseLevelH_SinglCell}) with (\ref{eq:effNoiseLevelH_SinglCell_TrainPhase}), we observe that the second term of
(\ref{eq:effNoiseLevelH_SinglCell}) is the gain due to data-aided channel estimation, and any data estimation error $\mse_{X_2} \neq 0$ results in a reduction of the gain.

The MSEs w.r.t.~$\qX_2$ under the pilot-only scheme and the JCD estimation scheme are shown in Figure \ref{fig:MSEX_All} as functions of the antenna ratio $\alpha$. The corresponding MSE of the Bayes-optimal estimator with perfect CSI is also plotted. As can be seen, the MSEs of all schemes decrease as $\alpha$ gets larger. Also, JCD estimation can improve the data estimation performance dramatically and approaches the perfect CSI case.

In addition, as $\beta$ becomes large, the JCD scheme is indistinguishable from the perfect CSI case. If $\beta$ is not so large, e.g., $\beta=5,10$, the gap does
not vanish even when the number of antennas becomes very large. To understand this, the MSEs w.r.t.~$\qH$ of the JCD scheme are also shown in Figure
\ref{fig:MSEX_All}. As $\alpha$ increases, the MSE w.r.t. $\qH$ does not vanish as expected and converges to a constant corresponding to $\mse_{X_2} = 0$ in
(\ref{eq:effNoiseLevelH_SinglCell}). Presumably, we may improve the channel estimation quality by increasing $\beta_1$. However, the improvement becomes marginal
in the JCD estimation scheme especially when the number of antennas is very large. Because $\beta$ is limited by the channel coherence time, increasing $\beta_1$
will decrease $\beta_2$ and thus reduce the data rate. This implies that negligible pilot information is preferable if JCD estimation is employed. This conclusion
is in agreement with \cite{Takeuchi-13TIT}, although a different receiver is considered. Note that negligible pilot information does not imply that no pilot
symbols are needed. As mentioned in Section II-B, pilot information (though almost negligible) is needed to avoid ambiguity problems.

We repeated the previous experiment with QPSK inputs and show the results in Figure \ref{fig:MSEXH_DataAid_QPSKANoiseless}. Unlike for Gaussian inputs, there are sharp transitions even for the comparatively high noise variance $\sigma^2 = 0.1$. If $\alpha$ exceeds the transition point, the MSE w.r.t.~$\qX_2$ approaches zero while the MSE w.r.t.~$\qH$ converges to a constant.

{\bf Example 3 (Two Cells and Unknown CSI)}---Consider a two-cell setup, i.e., $C=2$, with the model: $\qY = \qZ_1 + \qZ_2 + \qW$, where $\qZ_1 = \frac{1}{\sqrt{K}} \qH_1\qX_1 $ and $\qZ_2 = \frac{1}{\sqrt{K}}\qH_2\qX_2$ represent the signals from the desired users and the interference from neighboring cells, respectively. 
Since the pilot matrix $\qX_{1,1}$ is known, we substitute $\mse_{X_{1,1}} = 0$ into (\ref{eq:asyVarH}) and (\ref{eq:asyVarX}) to get
\begin{align}
 & \tq_{H_1} = \frac{\beta_1  \Gamma_{1}}{\sigma^2+ k_1 \Gamma_{1} \mse_{H_{1}} + k_2 \Delta_{2,1} } + \frac{\beta_2 (\Gamma_2-\mse_{X_{1,2}})}{\sigma^2 + k_1 \Delta_{1,2} + k_2 \Delta_{2,2} }, \label{eq:effNoiseLevelH_MultCell} \\
 & \tq_{X_{1,2}} = \frac{\alpha (G_{1}-\mse_{H_{1}})}{\sigma^2 + k_1 \Delta_{1,2} + k_2 \Delta_{2,2} }, \label{eq:effNoiseLevelX_MultCell}
\end{align}
where $\Delta_{c,t} = \mse_{H_c} \Gamma_t + \mse_{X_{c,t}} (G_c-\mse_{H_c})$.

For conventional JCD estimation \cite{Takeuchi-13TIT,Ma-14TSP}, the pilot information from the neighboring cells must be available at the target BS; otherwise, the interference from neighboring cells cannot be estimated and removed. Therefore, in the setting of our interest (in which $\qX_{2,1}$ is unavailable at the target BS), the interference $\qZ_2$ cannot be removed by conventional JCD estimation.

\begin{figure}
\begin{center}
\resizebox{3.5in}{!}{%
\includegraphics*{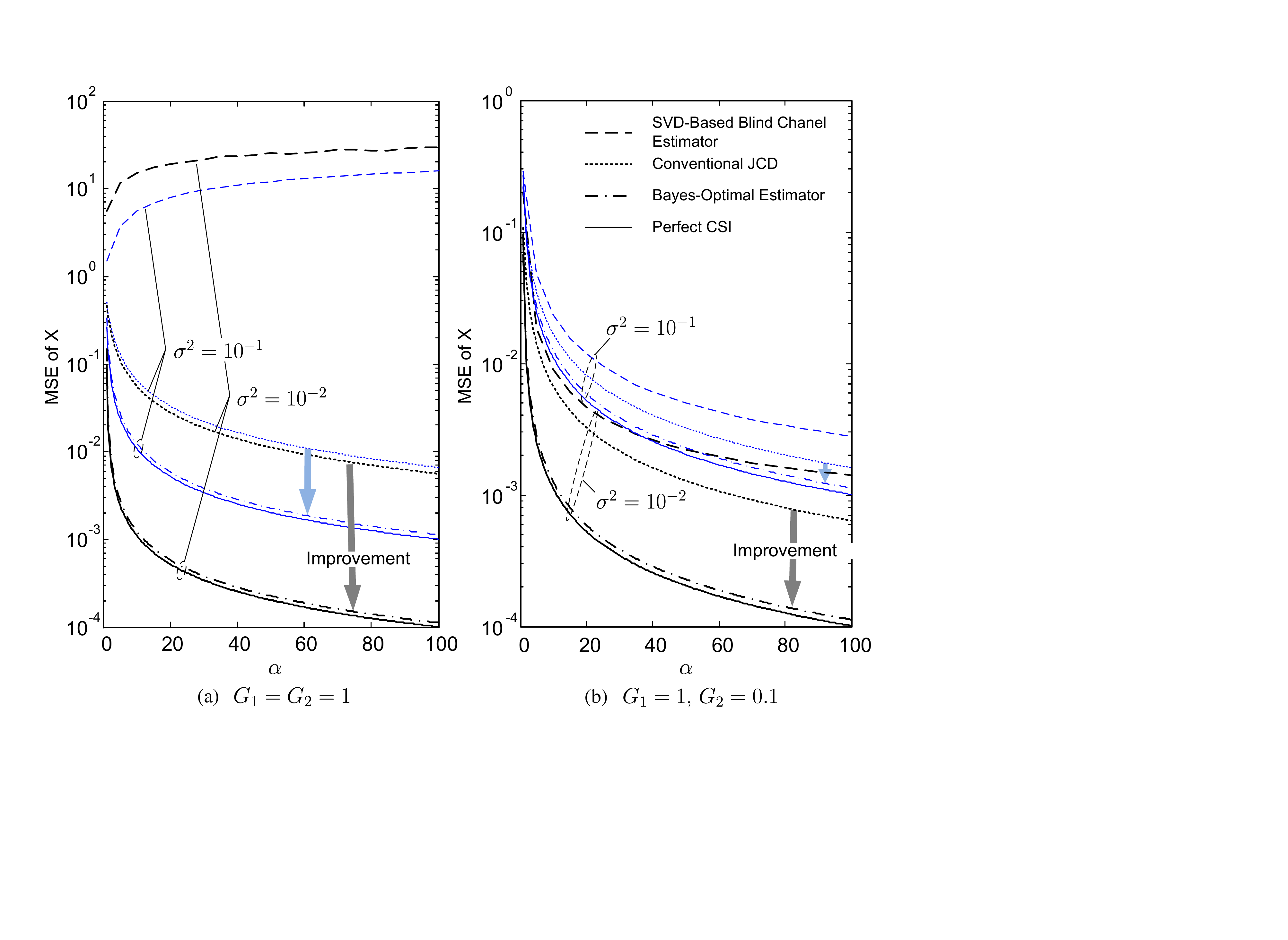} }%
\caption{The MSEs w.r.t.~$\qX_{1,2}$ as functions of $\alpha$ for different estimation schemes. $\beta=10$, $\beta_1=1$, $\Gamma_t=1,\,\forall t$, $H_{c}^{ij} \sim \calN_{\bbC}(0,G_c)$, and $X_{c,t}^{ij} \sim \calN_{\bbC}(0,\Gamma_t)$. }\label{fig:MSEXH_DataAid_Cell}
\end{center}
\end{figure}

For the results in Figure \ref{fig:MSEXH_DataAid_Cell}, we adopt the same parameters as were used for Figure \ref{fig:MSEX_All}. Here, we uniformly partition the
UEs into two groups: one group in the target cell and another group in the other cell, say Cell $2$. Two cases with 1) $G_1=G_2=1$ and 2) $G_1=1$ and $G_2=0.1$,
corresponding to, respectively, severe and mild interference scenarios, were considered. The MSEs of the SVD-based blind channel estimation scheme
\cite{Muller-14JSSP}, the conventional JCD estimation scheme \cite{Takeuchi-13TIT,Ma-14TSP}, the Bayes-optimal estimator, and the Bayes-optimal estimator with perfect CSI (for all the
cells) are provided. The MSEs of the SVD-based scheme were obtained by averaging over $10^4$ simulation trials, with $K=10$. We observe that the SVD-based scheme
performs the worst among the four schemes. It is noted that for the conventional JCD estimation scheme, we actually adopt the Bayes-optimal estimator while
completely ignoring $\qZ_2$, which results in $\Delta_{2,t} = G_2\Gamma_t, \forall t$. Therefore, the MSEs of the conventional JCD estimation scheme shown in Figure \ref{fig:MSEX_All} are expected to be better than those employing suboptimal criteria \cite{Takeuchi-13TIT,Ma-14TSP}. Even so, the Bayes-optimal estimator still shows a large
improvement over the conventional JCD estimation scheme. In fact, the MSEs w.r.t.~$\qX_{1,2}$ for the Bayes-optimal estimator are very close to the perfect CSI
case. Although not shown here, we find that the gap between the Bayes-optimal estimator and the perfect case can be reduced further by increasing $\beta$.

\section*{\sc VI. Conclusion}
Using the replica method, we have derived asymptotic expressions for the MSEs of the Bayes-optimal estimator for uplink transmission in a massive MIMO system. We found that with the Bayes-optimal estimator, the data symbols of the desired users can be reliably estimated even if only negligible pilot information is available
and the interfering signals from adjacent cells can be separated from the received signals without pilot information. Furthermore, the MSEs w.r.t.~the data symbols and the channels of the desired users decrease linearly with the numbers of antennas and data symbols, respectively. The large performance gaps between the Bayes-optimal estimator and the existing suboptimal estimators motives the search for low-complexity approximations of the optimal estimator and other improved suboptimal estimators for massive MIMO systems.

\section*{\sc Appendix}
We begin by rewriting $\Ex_{\qY}\{Z^{\tau}(\qY)\}$ using (\ref{eq:ProYcondHX}) and (\ref{eq:partFun}), which yields
\begin{equation} \label{eq:sf_E1}
    \Ex_{\qY}[Z^{\tau}(\qY)]
    = \Ex_{\calqH,\calqX}\left\{\int \rmd \qY \frac{\prod_{a=0}^{\tau} e^{-\frac{1}{\sigma^2}\left\| \qY - \frac{1}{\sqrt{K}}  \qH^{(a)} \qX^{(a)} \right\|^2}}{(\pi\sigma^2)^{NT}} \right\},
\end{equation}
where $\qH^{(a)}$ and $\qX^{(a)}$ are the $a$th replica of $\qH$ and $\qX$, respectively. For brevity, we define $\calqX \triangleq \{ \qX^{(a)}, \forall a\}$ and
$\calqH \triangleq \{ \qH^{(a)}, \forall a \}$. Note that $(\qH^{(a)},\qX^{(a)})$ are random matrices taken from the distribution $(P_H,P_X)$ for $a=0,1, \dots,
\tau$.

Next, we focus on calculating the right-hand side of (\ref{eq:sf_E1}), which can be done by applying the techniques in \cite{Krzakala-13ISIT,Kabashima-14ArXiv}
after additional manipulations. First, to carry out the expectations over matrices $\calqX$ and $\calqH$, we insert two identities which capture all combinations
of the replicas
\begin{align} 
1 &=  \int \prod_{n=1}^{N}\prod_{c=1}^{C} \prod_{0\leq a \leq b}^{\tau}\delta\left(
 \qh_{n,c}^{(b)} (\qh_{n,c}^{(a)})^{\dag} - K_c Q_{H_c}^{a,b} \right) \rmd Q_{H_c}^{a,b}, \\
1 &= \int \prod_{c=1}^{C}\prod_{t=1}^{2}\prod_{j \in \calT_t}\prod_{0\leq a \leq b}^{\tau}\delta\left(
 (\qx_{c,j}^{(a)})^{\dag} \qx_{c,j}^{(b)} - K_c Q_{X_{c,t}}^{a,b} \right)  \rmd Q_{X_{c,t}}^{a,b}, 
\end{align}
into (\ref{eq:sf_E1}), where $\delta(\cdot)$ denotes Dirac's delta. Here, $\qh_{n,c}^{(a)}$ denotes the $n$th row vector of $\qH^{(a)}$ corresponding to cell $c$,
and $\qx_{c,j}^{(a)}$ denotes the $j$th column vector of $\qX^{(a)}$ corresponding to cell $c$ and phase block $t$, and $\calT_t$ for $t=1,2$ represents the set of
all symbol indices in phase block $t$.

Let us define ${\calqQ_{H} \triangleq \{ \qQ_{H_c} = [Q_{H_c}^{a,b}] \in \bbC^{(\tau+1)\times(\tau+1)}, \forall c\}}$ and ${\calqQ_X \triangleq\{ \qQ_{X_{c,t}} =
[Q_{X_{c,t}}^{a,b}] \in \bbC^{(\tau+1)\times(\tau+1)}, \forall c,t \} }$. As a result, (\ref{eq:sf_E1}) can be rewritten as
\begin{equation}\label{eq:sf_E2}
 \Ex_{\qY}\{Z^{\tau}(\qY)\} = {\int e^{K^2{\cal G}^{(\tau)}}\rmd\mu_H^{(\tau)}(\calqQ_H) \rmd\mu_X^{(\tau)}(\calqQ_X)},
\end{equation}
where
\begin{equation}\label{eq:G1}
{\cal G}^{(\tau)} \triangleq \frac{1}{K^2} \log \Ex_{\calqZ}\left\{ \frac{1}{(\pi\sigma^2)^{NT}} \int \rmd\qY\prod_{a=0}^{\tau}e^{-\frac{1}{\sigma^2}
\left\|\qY - \qZ^{(a)} \right\|^2}\right\},
\end{equation}
and
\begin{align}
\mu_H^{(\tau)}(\calqQ_H) &\triangleq\Ex_{\calqH}\left\{\prod_{n,c,a,b} \delta\left(
 \qh_{n,c}^{(b)} ( \qh_{n,c}^{(a)} )^{\dag} - K_c Q_{H_c}^{a,b} \right)\right\},\\
\mu_X^{(\tau)}(\calqQ_X) &\triangleq\Ex_{\calqX}\left\{\prod_{c,t,j,a, b}^{\tau}\delta\left(
 (\qx_{c,j}^{(a)})^{\dag} \qx_{c,j}^{(b)} - K_c Q_{X_{c,t}}^{a,b} \right)\right\}.
\end{align}
In (\ref{eq:G1}), we have introduced random variables
\begin{equation} \label{eq:GaussAsum1}
    z_{n,j}^{(a)}\triangleq \frac{1}{\sqrt{K}} \qh_n^{(a)} \qx_{j}^{(a)},
    ~\mbox{for}~ a=0,1,\ldots,\tau,
\end{equation}
and ${\calqZ \triangleq \{ \qZ^{(a)} = [z_{n,j}^{(a)}] \in \bbC^{N\times T}, \forall a \}}$. The application of the central limit theorem suggests that the
$\qz_{n,j} \triangleq [ z_{n,j}^{(0)} \, z_{n,j}^{(1)} \cdots z_{n,j}^{(\tau)} ]^T$ are Gaussian random vectors with ${(\tau+1)\times(\tau+1)}$ covariance matrix
$\qQ_{Z_t}$. If $j \in \calT_t$, the $(a,b)$th entry of $\qQ_{Z_t}$ is given by
\begin{equation} \label{eq:defQ}
     (z_{n,j}^{(a)})^* z_{n,j}^{(b)}
     = \sum_{c=1}^{C} k_c Q_{H_c}^{a,b} Q_{X_{c,t}}^{a,b} \triangleq Q_{Z_t}^{a,b} .
\end{equation}
Because of the above-mentioned Gaussian property, we can calculate the expectation over $\calqZ$ after integrating over $\qY$ in (\ref{eq:G1}). Next, the remaining
integrals over $(\calqQ_H,\calqQ_X)$ can be evaluated via the saddle point method as $K \rightarrow\infty$ yielding $\Phi = \lim_{\tau\rightarrow
0}\frac{\partial}{\partial \tau}\Extr_{\calqQ_H,\calqQ_X,\tilde{\calqQ}_H,\tilde{\calqQ}_X} \{ \Phi^{(\tau)} \}$ with
\begin{align}\label{eq:saddlePoint1}
&\hspace{-0.2cm} \Phi^{(\tau)} \triangleq -\alpha \sum_{t} \beta_t\log \det\left( \qI + \qQ_{Z_t}\qSigma \right) - \alpha \beta \log\left( 1+\tau \right) + \notag \\
&\hspace{-0.2cm} \frac{1}{K^2} \log\Ex_{\calqH}\Big\{\prod_{n,c} e^{\tr\left(\tilde\qQ_{H_c}\qH_{n,c}^{\dag}\qH_{n,c}\right)}\Big\} -\sum_{c}\alpha k_c{\tr\left(\tilde\qQ_{H_c}\qQ_{H_c}\right)}+ \notag \\
&\hspace{-0.2cm} \frac{1}{K^2}  \log\Ex_{\calqX}\Big\{\prod_{c,t}e^{\tr\left(\tilde\qQ_{X_{c,t}}\qX_{c,t}^{\dag} \qX_{c,t}\right)}\Big\} -{\sum_{c,t}\beta_t k_c
 \tr\left(\tilde\qQ_{X_{c,t}}\qQ_{X_{c,t}}\right)},
\end{align}
where $\Extr_{x}\{ f(x) \}$ is the extreme value of $f(x)$ w.r.t.~$x$, $\qI$ denotes the identity matrix, $\tilde{\calqQ}_{H} \triangleq \{ \tilde{\qQ}_{H_c} =
[\tilde{\qQ}_{H_c}^{a,b}] \in \bbC^{(\tau+1)\times(\tau+1)}, \forall c\}$ and $\tilde{\calqQ}_X \triangleq\{ \tilde{\qQ}_{X_{c,t}} = [\tilde{\qQ}_{X_{c,t}}^{a,b}]
\in \bbC^{(\tau+1)\times(\tau+1)}, \forall c,t \} $.

The saddle points of (\ref{eq:saddlePoint1}) can be obtained by seeking the point of zero gradient
w.r.t.~$\{\qQ_{H_c},\qQ_{X_{c,t}},\tilde{\qQ}_{H_c},\tilde{\qQ}_{X_{c,t}}\}$. However, in doing so, it is prohibitive to get explicit expressions about the saddle
points. Therefore, we assume that the saddle points follow the {\em replica symmetry} (RS) form \cite{Nishimori-01BOOK} as $\qQ_{H_c} = (c_{H_c}-q_{H_c})\qI +
q_{H_c}\qone$, $\tilde\qQ_{H_c} =  (\tc_{H_c}-\tq_{H_c})\qI + \tq_{H_c}\qone$, $\qQ_{X_{c,t}} =(c_{X_{c,t}}-q_{X_{c,t}})\qI + q_{X_{c,t}}\qone$,
$\tilde\qQ_{X_{c,t}} = (\tc_{X_{c,t}}-\tq_{X_{c,t}})\qI + \tq_{X_{c,t}}\qone$, where $\qone$ denotes the all-one matrix. Also, we set $\qQ_{Z_t} = \sum_{c=1}^{C}
k_c ( (c_{H_c}c_{X_{c,t}}-q_{H_c}q_{X_{c,t}})\qI + q_{H_c}q_{X_{c,t}}\qone )$.

With the RS, we only have to determine the parameters $\{c_{H_c},q_{H_c},c_{{X_{c,t}}},q_{{X_{c,t}}},\tc_{H_c},\tq_{H_c},\tc_{{X_{c,t}}},\tq_{{X_{c,t}}} \}$, which
can be obtained by equating the corresponding partial derivatives of $\Phi^{(\tau)}$ to zero. Then, it is easy to check that $\tc_{H_c} =0$, $\tc_{{X_{c,t}}} = 0$, $c_{H_c} =
\Ex\{|{H_c}|^2\}$, and $c_{{X_{c,t}}} =\Ex\{|{X_{c,t}}|^2\}$. Let $\mse_{H_c} = c_{H_c}-q_{H_c}$ and $\mse_{{X_{c,t}}} = c_{{X_{c,t}}}-q_{{X_{c,t}}}$. We thus get
\begin{align}
\Phi&=-\alpha\sum_{t} \beta_t {\log\left( 1 + \sum_{c} \frac{k_c\Delta_{c,t} }{\sigma^2} \right)} -\alpha \beta \notag \\
&- \alpha \sum_{c} k_c I(H_c;Z_{H_c}|\tq_{H_c}) + \alpha \sum_{c} k_c \mse_{H_c}\tq_{H_c} \notag \\
&- \sum_{t,c} \beta_t k_c I(X_{c,t};Z_{X_{c,t}}|\tq_{X_{c,t}}) + \sum_{t,c} \beta_t k_c \mse_{X_{c,t}}\tq_{X_{c,t}},
\end{align}
where we have defined $\Delta_{c,t} \triangleq \mse_{H_c} c_{X_{c,t}} + \mse_{X_{c,t}} (c_{H_c}-\mse_{H_c})$, and the notation $I(A,Z_A|q_A)$ is used to denote the
mutual information between $A$ and $Z_A$ with $Z_A = \sqrt{q_A} A + W$ and $W \sim \calN_{\bbC}(0, 1)$. Finally, equating the partial derivatives of $\Phi$
w.r.t.~the parameters $\{\mse_{H_c}, \mse_{X_{c,t}},\tq_{H_c},\tq_{X_{c,t}} \}$ to zero gives (\ref{eq:asyMSE})--(\ref{eq:asyVarX}) in Proposition \ref{Pro_MSE}.

\renewcommand{\baselinestretch}{1.1}
\begin{footnotesize}

\end{footnotesize}

\end{document}